\documentclass[conference]{IEEEtran}
\IEEEoverridecommandlockouts
\usepackage{algorithmic}
\usepackage{amsmath,amssymb,amsfonts}
\usepackage{cite}
\usepackage{graphicx}
\usepackage{multirow}
\usepackage{textcomp}
\usepackage{xcolor}

\def\BibTeX{{\rm B\kern-.05em{\sc i\kern-.025em b}\kern-.08em
    T\kern-.1667em\lower.7ex\hbox{E}\kern-.125emX}}

\begin{document}

\makeatletter
\newcommand{\linebreakand}{
    \end{@IEEEauthorhalign}
    \hfill\mbox{}\par
    \mbox{}\hfill\begin{@IEEEauthorhalign}
}
\makeatother

\title{Development of a software complex for the diagnosis of dentoalveolar anomalies using neural networks}

\author{
    \IEEEauthorblockN{Alexander Kolsanov}
    \IEEEauthorblockA{
        \textit{Department of Operative Surgery and Clinical Anatomy} \\
        \textit{with a course of innovative technologies} \\
        \textit{FSBEI HE SamSMU MOH Russia} \\
        Samara, Russia \\
        avkolsanov@mail.ru
    }
    \and
    \IEEEauthorblockN{Nikolai Popov}
    \IEEEauthorblockA{
        \textit{Department of Pediatric Dentistry} \\
        \textit{FSBEI HE SamSMU MOH Russia} \\
        Samara, Russia \\
        2750668@mail.ru
    }
    \and
    \IEEEauthorblockN{Irina Aiupova}
    \IEEEauthorblockA{
        \textit{Department of Pediatric Dentistry} \\
        \textit{FSBEI HE SamSMU MOH Russia} \\
        Samara, Russia \\
        aupovaio@mail.ru
    }
    \and
    \IEEEauthorblockN{Konstantin Dobratulin}
    \IEEEauthorblockA{
        \textit{College of Information Technologies} \\
        \textit{and Computer Sciences} \\
        \textit{NUST MISIS} \\
        Moscow, Russia \\
        dobratulin@yahoo.com
    }
    \and
    \IEEEauthorblockN{Andrey Gaidel}
    \IEEEauthorblockA{
        \textit{Video Mining Laboratory} \\
        \textit{IPSI RAS} \\
        Samara, Russia \\
        andrey.gaidel@gmail.com
    }
}

\maketitle

\begin{abstract}
    This article describes the goals and objectives of developing a software
    complex for planning the treatment of dentoalveolar anomalies, the
    architecture of the software complex as interacting components for treatment
    planning, as well as the principle of using algorithms using convolutional
    neural networks within the software complex for a component that solves the
    problem of decoding a teleradiographic image.
\end{abstract}

\begin{IEEEkeywords}
    neural networks, computer vision, cephalometric analysis, teleradiography, orthodontics
\end{IEEEkeywords}

\section{Introduction} \label{Introduction}

The social significance of the diagnosis and treatment of dentoalveolar anomalies is due to their high prevalence, the tendency to an increase in the number of patients with this pathology, disorders from other organs and body systems associated with these pathologies \cite{c01}. The main research method in orthodontics for the diagnosis of anomalies and treatment planning is the cephalometric analysis of teleradiographic images in the lateral projection \cite{c02}. This analysis is a laborious and time-consuming research method that requires the proper experience and qualifications of a doctor \cite{c03}.

Using of neural networks affects the state of medicine - neural networks help doctors quickly and accurately decrypt images and reduce the number of medical errors \cite{c04}.
Neural networks have gained considerable popularity in applied problems of medicine as an auxiliary tool \cite{c05}.

Today it becomes possible to optimize the process using software technologies and algorithms approaches from the field of artificial intelligence.
The creation of software complex that would shorten the distance in the application of complex mathematical algorithms that underlie neural networks, and the end-user - a medical professional, is an urgent task that served as the purpose of developing the software described in this article.

\section{Neural networks in medicine} \label{Neural networks in medicine}

The use of neural networks in medicine and medical diagnostics began a long time ago, however, with the advent of new mathematical approaches and an increase in the power of personal computers of medical workers, it became possible to use more complex methods and algorithms \cite{c06}. So, thanks to a mathematical concept called convolutional neural networks, new solutions have become available for classification of images in medicine, segmentation of biomedical images, and so on \cite{c07, c08}. It was the concept of using convolutional neural networks in medicine that stimulated the creation of applied software that implements this concept and combines the capabilities of an intelligent image analysis algorithm and the simplicity of application software for interacting with the end-user.

\section{Subject area of software} \label{Subject area of software}

Research and diagnostics of patients in orthodontic treatment have become the subject area for the development of applied software \cite{c09}. There are many classifications of anomalies studied in orthodontics: jaw size anomalies, anomalies in the position of the jaws in the skull, anomalies in the ratio of dental arches, anomalies in the shape and size of dental arches, anomalies of individual teeth \cite{c10, c11}. An approach called cephalometric radiological image analysis, at the planning stage of treatment provides mathematical values for further analysis and diagnosis, such as jaw length and angles of inclination. The use of applied software based on the concept of using convolutional neural networks should significantly speed up the process of cephalometric analysis of radiological images, since the process itself is often performed manually or in a graphic editor without the ability to automatically obtain decoding \cite{c12}. The methodology for using convolutional neural networks has been studied earlier, therefore, at the software development step, we assign value to the parameters of the software being developed based on the neural network model \cite{c13}.

An approximate view of the result of decoding the image is shown in Fig.~\ref{image-analysis}.

\begin{figure}[htbp]
    \centerline{\includegraphics[scale=0.40]{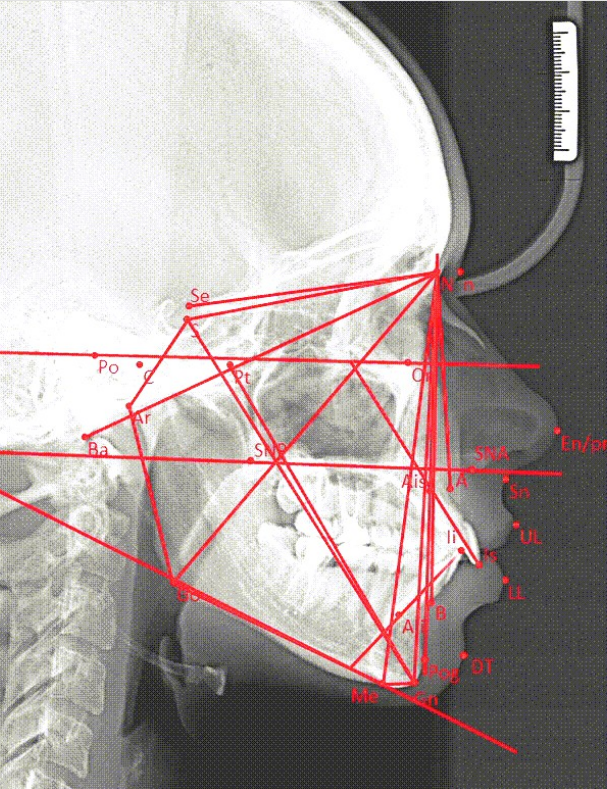}}
    \caption{Approximate view of the result of decoding the image.}
    \label{image-analysis}
\end{figure}

\section{Development of a software complex} \label{Development of a software complex}

\subsection{Technologies used}

We selected technologies for the development of a software complex:

\begin{enumerate}
    \item As the main programming language is Python 3 \cite{c14};
    \item Loading and preprocessing of images used the computer vision library named OpenCV \cite{c15};
    \item TensorFlow software library as one of the libraries providing implementation of computer vision algorithms and convolutional neural networks \cite{c16};
    \item Saving the result of localization of anatomical reference points is implemented using the following libraries: OpenCV to save the result of localization as an image; Pandas \cite{c17}, which provides powerful structures for storing and unloading data; Matplotlib library used for data visualization \cite{c18}.
\end{enumerate}

The connection scheme of the components from the stage of image appearance to the output of the results is shown in Fig.~\ref{image-scheme}.

\begin{figure}[htbp]
    \centerline{\includegraphics[scale=0.25]{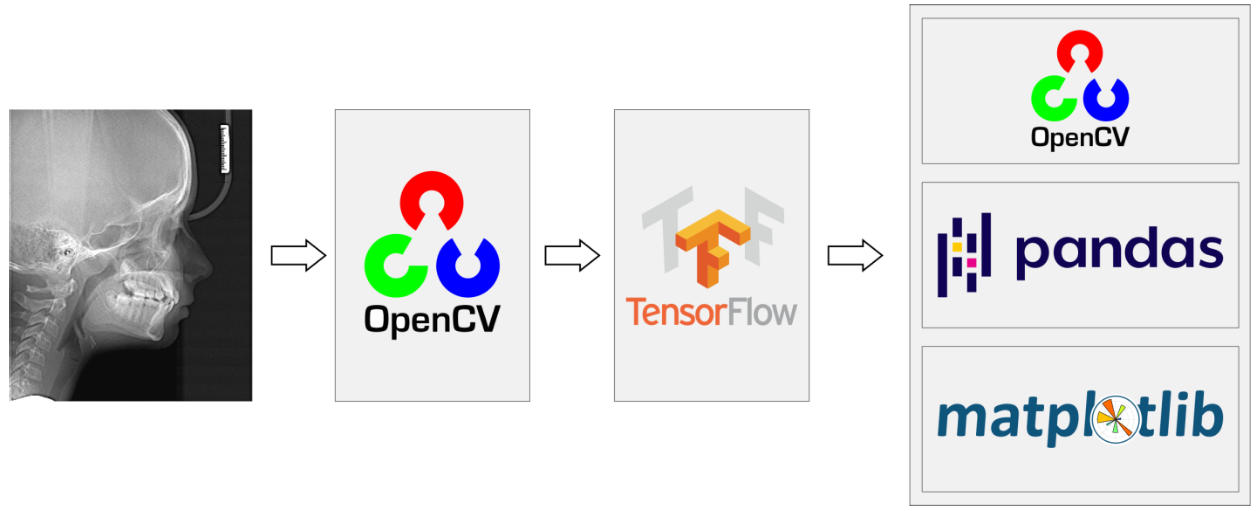}}
    \caption{The connection scheme of the components.}
    \label{image-scheme}
\end{figure}

\subsection{Interaction of software components and the process of decoding by a neural network}

The interaction of the components of the software package for decoding the teleradiographic image is reduced to the following process:

\begin{enumerate}
    \item The image is loaded from any possible JPEG / PNG / TIFF format using OpenCV;
    \item The primary image normalization is performed as \(X_{norm} = (X_{orig} - \min(X)) / (\max(X) - \min(X))\), where $X$ is a image matrix representation;
    \item The trained convolutional neural network is applied by means of TensorFlow, the obtained decoding results are extracted in the form of matrix representations;
    \item The resulting image decoding result is processed by the libraries: OpenCV to obtain a graphical interpretation of the result; Pandas to get a decoding report in a form understandable to the end-user; matplotlib for plotting metrics of decoding results.
\end{enumerate}

A preliminary visualization of the result of using a trained convolutional neural network using TensorFlow shown in Fig.~\ref{image-preresult}. Such a result, only more accurate according to the results of preliminary training of the model to the best parameters, is used for further analysis.

\begin{figure}[htbp!]
    \centerline{\includegraphics[scale=0.60]{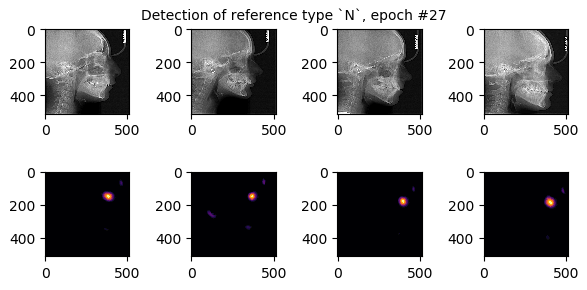}}
    \caption{A preliminary visualization of the result of using a trained convolutional neural network.}
    \label{image-preresult}
\end{figure}

\subsection{Development of a graphical interface for the end-user}

The interaction of the end-user with the software package should not be complicated and should not create obstacles before using neural network algorithms. The concept of an intuitive interface is the leading one in the design of a software package. At the first stage of development, for testing the software package on a pilot group of doctors to identify problems and assess the quality of the software package, we created a simple interface that provides image loading, automatic decoding using convolutional neural networks, and generating a report.

At the moment, the test interface with the ability to manually correct automatic decoding by moving the position of the marked reference points shown in Fig.~\ref{image-program}.
It should be noted that the presented interface is not the final version and serves to collect feedback and suggestions from the end-users of the test group, namely, a group of practicing orthodontists who agreed to test the developed software package.

\begin{figure*}[t!]
    \centerline{\includegraphics[width=\textwidth]{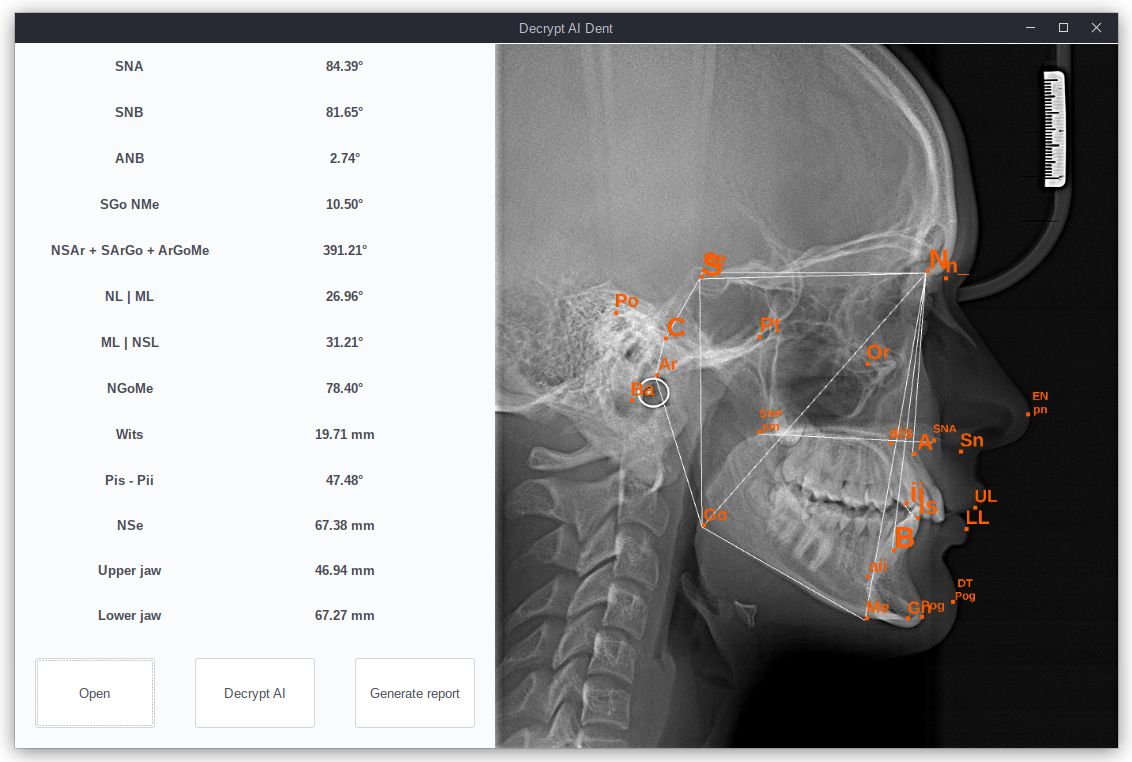}}
    \caption{The test interface of program.}
    \label{image-program}
\end{figure*}

\section{Results} \label{Results}

\subsection{Assessment of the applicability of the software package on individual PCs}

For testing on individual PCs of end-users, as well as for further commissioning of the software complex, we assess the characteristics obtained in the course of software development. So, the minimum requirements that we place on the software package are:

\begin{itemize}
    \item OS: Windows 7, 8, 10 x64, Ubuntu 18.04, 20.04
    \item Processor: 2 Ghz
    \item Memory: 2 GB RAM
    \item Storage: 4 GB available space
\end{itemize}

These requirements were formulated on the basis of an empirical assessment of the performance of the libraries used in this software package, as well as based on the space they occupy on the hard disk, taking into account the storage of the weights HDF5 file of the pre-trained convolutional neural network model, which takes about $110$ MB on hard drive \cite{c19}.

When decoding one and only one image, we do not require the end-user to have a GPU on his device, since the analysis of one image with the above requirements can be quite efficiently performed on the CPU. The time efficiency score ranges from $1$ second (Intel Core i5-8265U) to $3$ seconds (Intel Celeron N3450). Further testing in a focus group will allow us to provide the most accurate performance estimate when improving the software package.

\subsection{Potential integration with other software}

We are considering the potential for integrating the software package with existing systems used in the medical industry. In particular, the unification of the report format provided by the software package developed by us allows using the data for import into other software for the purpose of further research or building reports in a corporate form. For example, using the Pandas library at the stage of report generation allows you to create reports in CSV format, which is extremely common for storing and analyzing tabular data \cite{c20}.

\section{Summary} \label{Summary}

As a result of the development of a software package based on the use of algorithms using convolutional neural networks, architectural solutions are described and estimates of the parameters of the development and operation cycle are obtained, such as the average speed of program execution when using an algorithm using a convolutional neural network for the analyzed image; the size of the parameter file used in the software package, which makes it possible to conclude that the development of the software package for operation on personal computers is promising. The results obtained are indicative when integrating the software package into an industrial environment.

\vspace{12pt}

\end{document}